\documentclass[3p,twocolumn,sort&compress,showpacs,showkeys,preprintnumbers,amsmath,amssymb,aps,floatfix,pre]{elsarticle}

\usepackage{lineno}
\modulolinenumbers[5]

\usepackage{graphicx}
\usepackage{epstopdf}
\usepackage{amsmath}
\usepackage{multirow}
\usepackage{times}
\usepackage{amssymb}
\usepackage{dcolumn}%
\usepackage{bm}%
\usepackage[dvipdfmx]{hyperref}%
\usepackage{upgreek}
\usepackage{wasysym}
\usepackage{verbatim}

\hypersetup
{
	colorlinks=true,
	linkcolor=blue,
	citecolor=blue
}

\journal{Journal of \LaTeX\ Templates}

\begin{document}

\begin{frontmatter}

\title{Bond disorder, frustration and polymorphism \\ in the spontaneous crystallization of a polymer melt}

\author{A. Giuntoli}
\address{Dipartimento di Fisica ``Enrico Fermi'', 
Universit\`a di Pisa, Largo B.\@Pontecorvo 3, I-56127 Pisa, Italy}


\author{S. Bernini}
\address{Dipartimento di Fisica ``Enrico Fermi'', 
Universit\`a di Pisa, Largo B.\@Pontecorvo 3, I-56127 Pisa, Italy}
\fntext[myfootnote]{present address: Jawaharlal Nehru Center for Advanced Scientific Research, Theoretical Sciences Unit, Jakkur Campus, Bengaluru 560064, India.}

\author{D. Leporini*}
\address{Dipartimento di Fisica ``Enrico Fermi'', 
Universit\`a di Pisa, Largo B.\@Pontecorvo 3, I-56127 Pisa, Italy}
\address{IPCF-CNR, UOS Pisa, Italy}

\cortext[D. Leporini]{Corresponding author}
\ead{dino.leporini@unipi.it}

\begin{abstract}
The isothermal, isobaric spontaneous crystallization of a supercooled polymer melt is investigated by MD simulation of an ensemble of fully-flexible linear chains.
Frustration is introduced via two incommensurate length scales set by the bond length and the position of the minimum of the non-bonding potential. 
Marked polymorphism with considerable bond disorder, distortions of both the local packing and the global monomer arrangements is observed. 
The analyses in terms of: i) orientational order parameters characterizing the global and the local order and ii) the angular distribution of the next-nearest neighbors of a monomer
reach the conclusion that the polymorphs are arranged in distorted Bcc-like lattices.
\end{abstract}


\end{frontmatter}

\section{Introduction}

Crystallization plays an important role in many areas of different scientific fields, ranging from 
biology to engineering and physics. Still, many microscopic details of the phenomenon are unknown, despite the abundance of related results both experimental and theoretical 
\cite{PhillipsCry90,SumpterCry90,KarayiannisIJMS13AthHSC,RussoTanaka12,LeomachTanaka12,MRJonesDNA10,NykypanchukDNA08,Gundlach08,Campoy-Quiles08,Hegedus08,Loo01}. 
In particular, polymeric liquids are systems in which the structural features, namely the chain connectivity, cause serious hindrance to the homogeneous 
crystallization of the sample. In silico simulations of such systems provide great insight on this problem and have proven to be an invaluable tool in the analysis of the
crystallization under controlled conditions \cite{KarayiannisIJMS13AthHSC}. 
Many simulations have been performed aiming at observing the crystallization of polymers and characterizing the structural order reached by the crystalline sample under 
various conditions \cite{Tanemura77,HahnCondis,TrappeJamming01,MeyerFold02,StachurskiIdealSolid03,AnikeenkoPacking08,IkedaJamming12,NiBondFluct12,LopatinaGranular14}.
Recently, Monte Carlo (MC) simulations of polymer melts 
made by linear chains of tangent hard-sphere monomers \cite{KarayiannisCEE09,KarayiannisPRL09HSC,KarayiannisSM10AthHSC}, i.e. with bond length {\it equal} to monomer diameter, 
have been performed to study spontaneous crystallization.
The resulting crystallized structures have been interpreted as a distribution of the most densely packed
structures: face-centered cubic (Fcc) and hexagonal close packed (Hcp) lattices. Hcp and Fcc were selected as ideally ordered structures because they are known to be the primary competing
alternatives in dense systems of hard spheres  in the presence of a single length scale \cite{KarayiannisCEE09}.
MC simulations, differently from  Molecular Dynamics (MD) simulations, may fail to account for the arrest into metastable intermediate phases
\cite{KarayiannisPRL09HSC} which hinder the evolution towards the thermodynamically stable phase \cite{WoldeOstwald99}. 
MD simulations of a polymer melt of chains with soft monomers,
promoting the crystallization by equal bond length and equilibrium non-bonded separation, have been performed with the aim of comparing the crystalline structures
obtained by cooling down to zero temperature with the highly packed Fcc and Hcp lattices \cite{KarayiannisPRE13Quench,KarayiannisBendingAngle15}.
Still, the route towards the closest packing of polymers is hindered  by allowing length-scale competition  of  the bonding and the 
non-bonding interactions, as recently proven in a MD study of the crystallization triggered by confinement due to Fcc walls, where structures similar to body-centered cubic (Bcc) are 
observed \cite{SimmonsBcc13}. 

Polymorphism, the presence of different crystal structures of the same molecule, is a well-known phenomenon in molecular crystals \cite{PolymorphismBernstein}.
In particle systems the crystal structure depends on the steepness of the repulsive part of the interacting potential  with hard and soft repulsions favoring Fcc and 
Bcc ordering respectively \cite{MilsteinPRB70}. To date, the selection mechanism of polymorphs is elusive. One widely used criterion is the Ostwald step rule, stating that 
in the course of transformation of an unstable,  or metastable  state, into a stable one the system does not go directly to the most stable conformation but prefers to reach intermediate stages
having the closest free energy to the initial state \cite{Ostwald1897,WoldeOstwald99,LariniCrystJCP05,PolymorphismBernstein}. Alternatives are reported \cite{TanakaRussoPolymorphSelSoftMatter12}.

In this work the isothermal {\it spontaneous} crystallization of an unbounded polymeric system is studied via MD simulation of {\it fully-flexible} linear chains, i.e. bond-bending and bond-torsions 
potentials are not present. The emphasis  is on the global and local order of the crystalline phase with respect to the pristine supercooled liquid where crystallization started. 
To this aim, specific order parameters will be used for their characterization \cite{Steinhardt83,LocalOrderJCP13}.
A distinctive feature of the model is the presence of two {\it different} length scales, namely the bond length $b$ and the distance $\sigma^*$ where the minimum of the non-bonding potential,
the Lennard-Jones (LJ) pair potential, is located. It is known that the competition of two incommensurate length scales favors frustration in the self-assembly of ordered structures from an 
initial disordered state, like in molecular crystallization \cite{Hamley2007}. Frustrated crystallization of polymers has been reviewed  \cite{WangChemRev03}. 
The role played in the crystallization behavior (including its absence) by frustration, where there is an
incompatibility between the preferred local order and the global crystalline order, has been highlighted \cite{DoyeFrustratCrystalPhysChemChemPhys07}.
We expect different responses to frustration from the putative crystalline structures at finite temperature, i.e. Fcc, Hcp and Bcc lattices. 
In fact, not all the atoms in the first neighbors shell of a Bcc lattice are at the same distance, as in the Fcc and Hcp lattices. 
It is known that the mechanical stability of the Bcc structure is lower than in closed packed structures as Fcc \cite{MilsteinPRB70}.

The paper is organized as follows: In Sec.\ref{numerical} the polymer model is detailed and the simulation details are provided.
The results are presented and discussed in Sec.\ref{resultsdiscussion}. Finally, the conclusions are drawn in Sec.\ref{conclusions}. 

\section{Methods}
\label{numerical}

We consider a coarse-grained polymer model of $N_c=50$ linear, unentangled chains with $M=10$ monomers per chain. The total number of monomers is $N=500$.  The chains are fully-flexible,
i.e. bond-bending and bond-torsions potentials are not present. Non-bonded monomers at distance $r$  interact via the truncated Lennard-Jones (LJ) potential: 
\begin{equation}
U^{LJ}(r)=\varepsilon\left [ \left (\frac{\sigma^*}{r}\right)^{12 } - 2\left (\frac{\sigma^*}{r}\right)^6 \right]+U_{cut}
\end{equation}
for $r\leq r_c=2.5\,\sigma$ and zero otherwise, where $\sigma^*=2^{1/6}\sigma$ is the position of the potential minimum with depth $\varepsilon$. The value of the constant
$U_{cut}$ is chosen to ensure that $U^{LJ}(r)$ is continuous at $r = r_c$. Henceforth, all quantities are expressed in terms of reduced units:  lengths in units of $\sigma$,
temperatures in units of $\varepsilon/k_B$ (with $k_B$ the Boltzmann constant) and time $\tau_{MD}$ in units of $\sigma \sqrt{m / \varepsilon}$ where 
$m$ is the monomer mass. We set $m = k_B = 1$.  The bonding interaction is described by an harmonic potential $U^b$ \cite{MauriLepEPL06}:
\begin{equation}
 U^b(r)=k(r-r_0)^2
\end{equation}
The parameters $k$ and $r_0$
have been set to $2500 \, \varepsilon  / \sigma^2 $ and $ 0.97\,\sigma $ respectively \cite{GrestPRA33}. Given the high stiffness of the bonding interaction, $b = 0.97 \pm 0.02$.
Notice that the bond length and the minimum of the non-bonding potential are {\it different}, $b \neq \sigma^* \simeq 1.12$.  Periodic boundary conditions are used. The study was performed
in the $NPT$ ensemble (constant number of particles, pressure and temperature). The integration time step is set to $\Delta t=0.003$ time units \cite{Puosi11,UnivPhilMag11,BarbieriGoriniPRE04,AlessiEtAl01,leporiniPRA94,LeporiniJPCM99}.
The simulations were carried out using LAMMPS molecular dynamics software (http://lammps.sandia.gov) \cite{PlimptonLAMMPS}. The samples were initially equilibrated at temperature $T=0.7$ 
and pressure $P=4.7$ for a lapse of time which is, at least, three times the average reorientation time 
of the end-end vector of the chain \cite{DoiEdwards}. After equilibration, we started production runs waiting for the spontaneous crystallization of the system.
We analyzed the initial equilibrated liquid state, the development of the solid phase and the final crystalline state. 56 starting configurations of the liquid were used with different 
random velocities, chain conformations and non-overlapping monomer positions. 42 equilibrated runs underwent crystallization, while 14 of them failed to crystallize in a reasonable amount of time (about one month of computing time).

\begin{figure}[t]
\begin{center}
\includegraphics[width=0.95\linewidth]{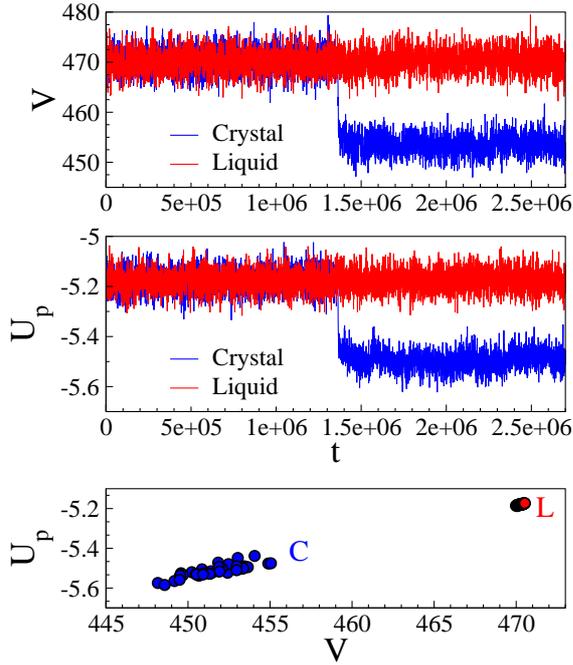}
\end{center}
\caption{Volume (top) and potential energy (middle) drops due to the spontaneous crystallization occurring in a single run (blue curve). 
They are  compared to the typical fluctuations occurring in the metastable liquid (red curve). The bottom panel is a correlation plot between the average volume and the energy 
of all the crystalline (blue) and liquid (red) states under study. Note the large region spanned by the different crystalline states signaling polymorphism.}
\label{Fig1}
\end{figure}

\begin{figure}[t]
\begin{center}
\includegraphics[width=0.95\linewidth]{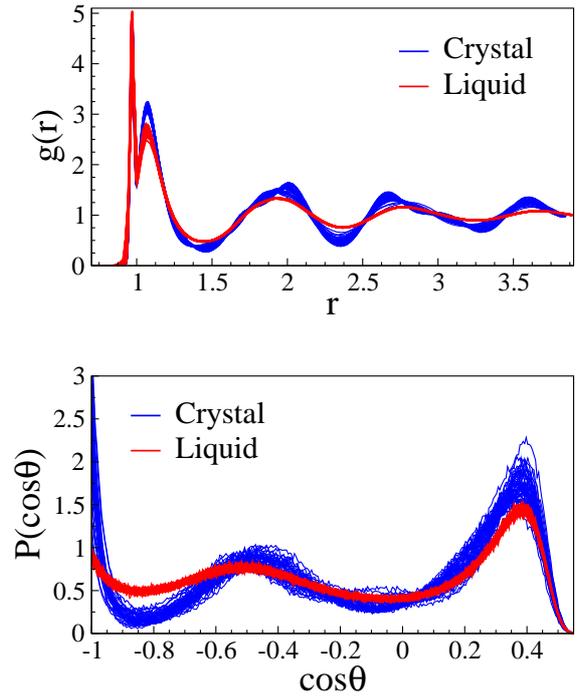}
\end{center}
\caption{Top: radial pair distribution function $g(r)$ of all the crystalline and liquid states. On increasing the distance $r$ from the tagged central monomer, the first, sharp peak corresponds
to the bonded monomers, whereas the other ones signal the different neighbor shells. Bottom: distribution $P(\cos \theta)$ of the angle $\theta$ between adjacent bonds in a chain.
The peaks occur at $\theta \approx 70^\circ, 122^\circ, 180^\circ$, corresponding to three consecutive monomers which are folded - with the two non-consecutive monomers in contact  ($r \sim\sigma^*$) -,
partially folded, and aligned, respectively \cite{LocalOrderJCP13}. 
Notice that the broad features of both $g(r)$ and $P(cos\theta)$ of the crystalline and the melt states are quite similar.
Nonetheless, the ordered states exhibit sharper and, due to polymorphism, more widely distributed features than the disordered ones.}
\label{Fig2}
\end{figure}

\section{Results and discussion}
\label{resultsdiscussion}

The spontaneous crystallization of the polymeric system is characterized by a sudden drop of both the volume $V$ and the potential energy $U_p$ during 
the time evolution of the system, as shown in Fig.\ref{Fig1}. It is seen that the mean values of both the volume and the potential energy of the crystals span a wider range 
with respect to the the metastable liquid, see lower panel of Fig.\ref{Fig1}. This is evidence of polymorphism due to different kinetic pathways leading to crystallization in more than one,
metastable, ordered form \cite{PolymorphismBernstein}. In our polymer melt polymorphism is contributed, with respect to the corresponding - non-bonded - atomic liquid, by the chain 
connectivity and the presence of incommensurate length scales involving the bonding and the non-bonding potentials.

In order to start the characterization of the polymorphism, we consider the arrangements of both the monomer and the bonds by the radial distribution function $g(r)$ and the distribution 
$P(\cos \theta)$ of the angle $\theta$ between adjacent bonds in a chain, respectively.  The results are reported in Fig.\ref{Fig2} and are compared to  the corresponding ones of the liquid.
It is apparent that $g(r)$ and $P(\cos \theta)$ are not markedly different in the polymorph and the liquid states even if
the polymorphs exhibit sharper, and more widely distributed, features, including a larger fraction of aligned bonds. 

\begin{figure}[t]
\begin{center}
\includegraphics[width=0.95\linewidth]{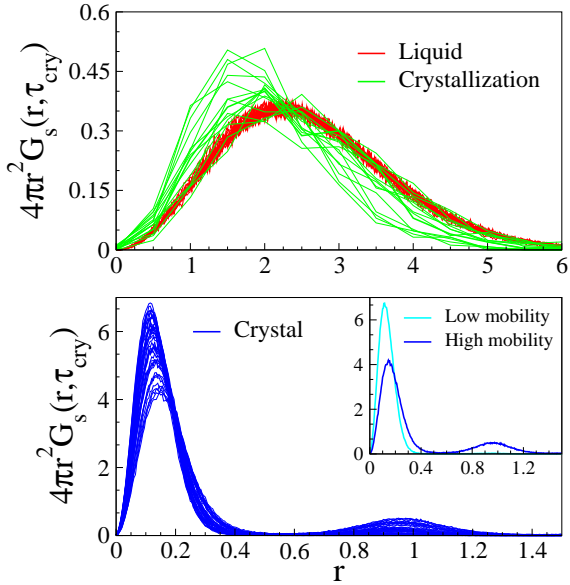}
\end{center}
\caption{Displacement distribution of the monomers in a time $\tau_{cry}$ during different stages of the crystallization. 
The average time to start and complete the crystallization is $\tau_{cry} = 12000$. 
Top: initial supercooled liquids in metastable equilibrium (red), selected supercooled liquid states undergoing crystallization (green). Different curves with same color refer to different crystallization paths. 
Bottom: final polymorphs. The small peak at $r\sim1$ signals the presence of monomers jumping of about one diameter. 
The initial liquid state of the paths exhibits always nearly the same displacement distribution.
The displacement distribution during the crystallization process is only mildly narrower than the one of the melt. 
The different polymorphs have distinct displacement distributions and lower mobility than the liquid. The inset of the bottom panel shows the two polymorphs with highest and lowest 
mobility, respectively. The jump process is largely suppressed in the less mobile polymorph.}
\label{Fig3} 
\end{figure}

The description of the selection mechanism of the polymorph is beyond the purpose of the present work. As initial step, we study the monomer mobility during the transition from the liquid 
to the polymorph.
To this aim, we resort to the self-part of the van Hove function $G_s(r,t)$ \cite{HansenMcDonaldIIIEd}. The product $G_{s}(r,t) \cdot 4\pi r^{2}$ is the probability that the monomer
is at a distance between $r$ and $r+dr$ from  the initial position after a time $t$.
We observe that the crystallization completes in a range of times spanning from about $ 0.8 \cdot 10^4$ up to $ 1.5 \cdot 10^4$ MD times for the different runs. The average value of the crystallization time is $\tau_{cry} = (1.2\pm0.1) \cdot 10^4$ time units, corresponding to $22\pm2$ ns mapping the MD units on polyethylene, according to the procedure outlined in ref. \cite{Kroger04}. Notice that the MD-polyethylene conversion factor, $\sigma \sqrt{m/ \varepsilon} = 1.8$ ps, is in the range usually found for polymers, i.e. $ \sim 1-10$ ps/ MD time step \cite{Kroger04}. Our average crystallization time is intermediate between the induction time, i.e. the average time to reach the critical nucleus size and proceed with further growth at later times \cite{WedekindMDActProcJCP07}, of $n$-octane, $16 \pm 10$ ns \cite{MDOctaneRutledgeJCP09},  and {\it n}-eicosane , $80.6 \pm 8.8$ ns 
\cite{MDC20RutledgeJCP09}, as evaluated by using a realistic, united-atom MD model for
{\it n}-alkanes.
No clear correlation between the crystallization time and crystal type was observed.
For the different crystallization paths we evaluate the displacement distribution in a time $\tau_{cry}$ of: i) the initial liquid in metastable equilibrium, 
ii) the liquid during crystallization and iii) the final polymorph.
The results are in Fig.\ref{Fig3}. During the crystallization monomers displace nearly as far as in the liquid. 
Starting from liquid states with nearly identical displacement distribution, see Fig.\ref{Fig3} (top), 
each subsequent crystallization path and final polymorph is characterized by a different displacement distribution. Polymorphs have little mobility. 
Their displacement distribution exhibits a bimodal structure with a large peak corresponding to the rattling motion of the monomer within the cage of the first neighbors
and a secondary peak due to monomers displacing by jumps with size comparable to their diameter. 
The inset of Fig.\ref{Fig3}  shows the  displacement distribution of the polymorphs with the highest and the lowest mobility and evidences that the 
jump process is largely suppressed in the less mobile polymorph.

\begin{figure}[t]
\begin{center}
\includegraphics[width=0.8\linewidth]{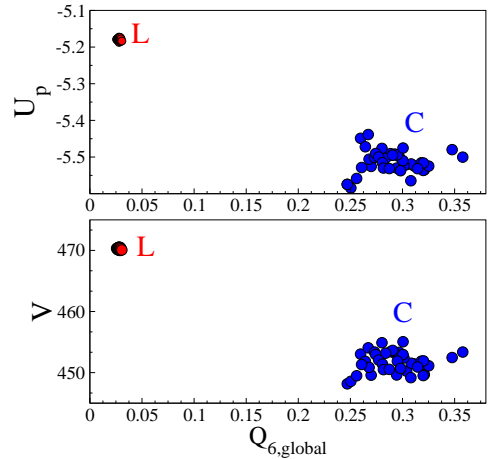}
\end{center}
\caption{Correlation plot between the global order parameter $Q_{6,global}$ and both the potential energy (top) and volume (bottom). Both liquid states (red) and polymorphs (blue) are shown. 
Polymorphs span an extended region of the plot with larger  global order than the liquid.}
\label{Fig4}
\end{figure}

\begin{figure}[t]
\begin{center}
\includegraphics[width=0.95\linewidth]{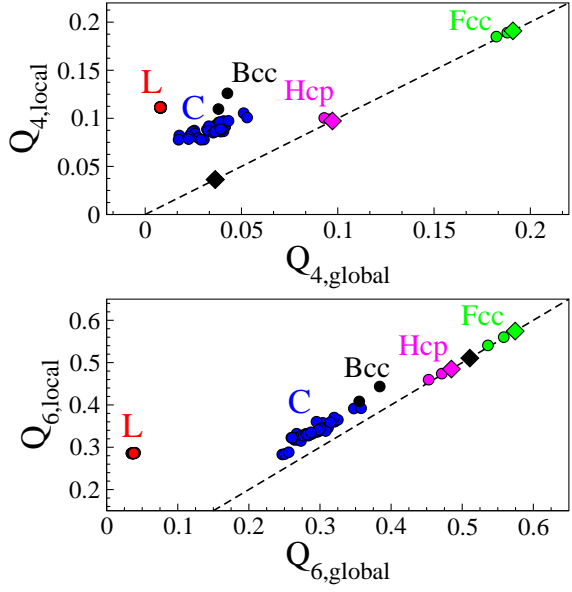}
\end{center}
\caption{Correlation plots between the local and the global parameters with $l = 4$ (top) and $l=6$ (bottom) for polymorphs (blue dots) and liquids (red dots). The dashed line is the bisectrix.
The diamonds mark the ideal Bcc- (black), Fcc- (green) and Hcp- (magenta) atomic crystals at T=0 with $Q_{l,global}=Q_{l,local}$. The black, green and magenta dots mark the same crystals at $T=0.3,0.7$.
Differently from the Bcc lattice, the global and the local order of the Fcc and Hcp lattices are negligibly affected by the temperature, signaling more thermal stability. 
In each plot the black dot closest to the polymorph region is the Bcc lattice at the same temperature ($T=0.7$).}
\label{Fig5}
\end{figure}

In order to study more rigorously the structural order of the system, we resort to the order parameters defined by Steinhardt \textit{et al.} \cite{Steinhardt83}.
One considers in a given coordinate system the polar and azimuthal angles $\theta({\bf r}_{ij})$ and $\phi({\bf r}_{ij})$ of the
vector ${\bf r}_{ij}$ joining the $i$-th central monomer with the $j$-th one belonging to the neighbors within a
preset cutoff distance $r_{cut} = 1.2 \; \sigma^* \simeq 1.35$ \cite{Steinhardt83}. $r_{cut}$ is a convenient definition of
the first coordination shell size \cite{sim}. The vector ${\bf r}_{ij}$ is usually referred to as a ``bond'' and has
not to be confused with the {\it actual} chemical bonds of the polymeric chain.
To define a global measure of the order in the system, one then introduces the quantity:
\begin{equation} \label{Qbarlm_global}
	\bar{Q}_{lm}^{global}=\frac{1}{N_{b}}\sum_{i=1}^{N}
\sum_{j=1}^{n_b(i)}Y_{lm}\left[\theta({\bf
r}_{ij}),\phi({\bf r}_{ij})\right]
\end{equation}
where $n_b(i)$ is the number of bonds of $i$-th particle, $N$ is the total number of particles in the system, $Y_{lm}$
denotes a spherical harmonic and $N_b$ is the total number of bonds:
\begin{equation} \label{N_b}
	N_b=\sum_{i=1}^{N} n_b(i) 
\end{equation}
The global orientational order parameter $Q_{l,global}$ is defined by:
\begin{equation} \label{Ql_global}
 Q_{l,global}=\left [ \frac{4\pi}{(2l+1)} \sum_{m=-l}^{l}
|\bar{Q}_{lm}^{global}|^2 \right ]^{1/2}
\end{equation}
The above quantity is invariant under rotations of the coordinate system and takes characteristic values which can be used to quantify the kind and the degree of
rotational symmetry in the system \cite{Steinhardt83}. In the absence of {\it large}-scale order, the bond orientation is uniformly distributed around the unit sphere and $Q_{l,global}$ 
is rather small since it vanishes as $ \sim N_b^{-1/2}$ \cite{RintoulTorquatoJCP96}. On the other hand, $Q_{6,global}$ is very sensitive to any kind of crystallization and increases significantly
when order appears \cite{GervoisGeometrical99}.
A local orientational parameter $Q_{l,local}$ can also be defined. We define the auxiliary quantity
\begin{equation} \label{Qbarlm_local}
	\bar{Q}_{lm}^{local}( i )=\frac{1}{n_b(i)}\sum_
{j=1}^{n_b(i)}Y_{lm}\left[\theta({\bf r}_{ij}),\phi({\bf
r}_{ij})\right]
\end{equation}
The local order parameter $Q_{l,local}$ is defined as \cite{Steinhardt83}:
\begin{equation} \label{Ql_local}
 Q_{l,local}=\frac{1}{N} \sum_{i=1}^{N}  \left [
\frac{4\pi}{(2l+1)} \sum_{m=-l}^{l} |\bar{Q}_{lm}^{local}( i
)|^2 \right ]^{1/2}
\end{equation}
In general $Q_{l,local}\ge Q_{l,global}$. In the presence of ideal order, {\it all} the particles have the {\it same} neighborhood configuration, and the equality $Q_{l,local} = Q_{l,global} = Q_l$
follows.

\begin{figure}[t]
\begin{center}
\includegraphics[width=0.95\linewidth]{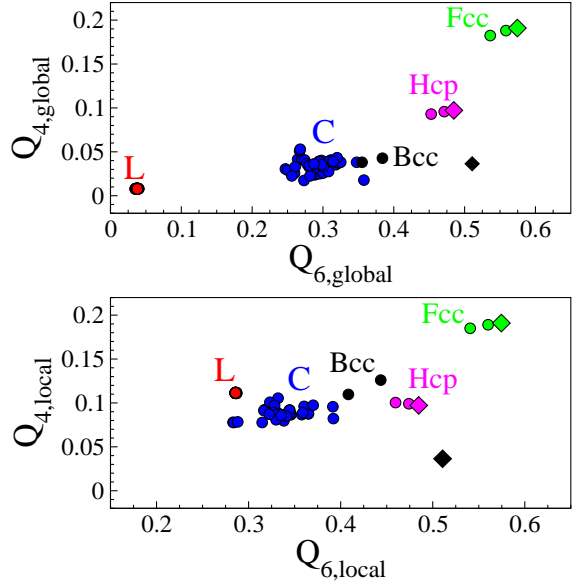}
\end{center}
\caption{Correlation plots  $Q_{4,global}$ vs. $Q_{6,global}$ (top) and $Q_{4,local}$ vs. $Q_{6,local}$ (bottom). All the symbols are used as in Fig.\ref{Fig5}. In each plot the black dot closest to the polymorph region is the Bcc lattice at the same temperature ($T=0.7$).}
\label{Fig5bis}
\end{figure}

\begin{figure}[t]
\begin{center}
\includegraphics[width=0.95\linewidth]{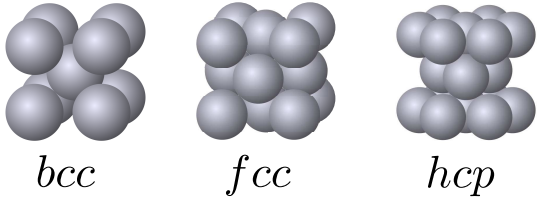}
\end{center}
\caption{Bcc-, Fcc- and Hcp-lattices at $T=0$. Fcc and Hcp structures are closely packed and each atom is surrounded by twelve neighbors at the distance $a$. 
In the Bcc structure each atom has neighbors at two different distances: $a$ and $2a/\sqrt{3}$.}
\label{Lattice}
\end{figure}

Fig.\ref{Fig4} shows the correlations between the volume $V$, the potential energy $U_p$ and the global order parameter $Q_{6,global}$ for both the liquids and the polymorphs.
We see that liquid states have rather small $Q_{6,global}$ with narrow distributions of $V$, $U_p$ and $Q_{6,global}$. The polymorphs have much larger $Q_{6,global}$ values with wider distributions
of $V$, $U_p$ and $Q_{6,global}$.

\begin{figure}[t]
\begin{center}
\includegraphics[width=0.95\linewidth]{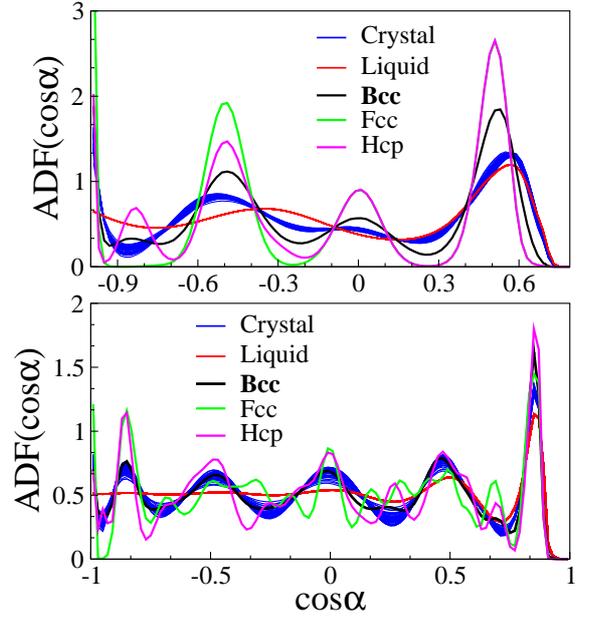}
\end{center}
\caption{Angular distribution function (ADF) in the first ($0.8<r<1.35$, top panel) and second ($1.35<r<2.2$, bottom panel) shells of the liquids and the polymorphs.
As reference, the distributions of Bcc, Fcc and Hcp crystals at $T=0.7$ are also plotted. Different polymorphs exhibit distinct ADFs.
No clear conclusion is drawn by comparing the ADF of the crystalline polymers and the ones of the atomic Bcc, Fcc and Hcp  crystals in the first shell. 
Instead, the ADF of the crystalline polymers and the Bcc crystal at $T=0.7$ is almost identical in the second shell.}
\label{Fig6}
\end{figure}

An interesting features of the order parameters is that they exhibit specific values for the different kind of orders \cite{Steinhardt83}. This suggests to plot their values to gain insight
into the global and the local order of the polymorphs and the liquid. This is done in Fig.\ref{Fig5}  and Fig.\ref{Fig5bis}. 
First, let us refer to Fig.\ref{Fig5} providing correlation plots $Q_{l,global}$ vs $Q_{l,local}$ for $l=4,6$. We see that $Q_{l,global}$ is higher in the polymorphs than in the liquids,
whereas $Q_{l,local}$ is nearly the same. In order to better understand the results, we added to Fig.\ref{Fig5}  the  Bcc, Fcc and Hcp crystals at different temperatures as reference crystals.
In the reference crystals atoms interact via the LJ potential.  At $T=0$ the closest particles are spaced by the average monomer-monomer distance of the polymer system, $a=1.07$. Their structure is 
shown in Fig.\ref{Lattice}. Bcc, Fcc and Hcp crystals at $T=0$ are characterized by the equality $Q_{l,global}=Q_{l,local}$. We also included the same atomic crystals at $T=0.3,0.7$. Fig.\ref{Fig5} shows that, when increasing the temperature, the order parameters of both Fcc and Hcp atomic lattices show little changes, whereas the ones of the Bcc crystal are much more sensitive. 
This is consistent with the known lower stability of the Bcc crystal \cite{MilsteinPRB70}. Fig.\ref{Fig5}, especially the case with $l=6$ (bottom panel), shows that the polymorphs position
themselves in an region close to the one of the Bcc structure at the same temperature ($T=0.7$). This is a piece of evidence that our melt of fully-flexible chains crystallizes spontaneously into a Bcc-like
structure distorted by the polymer connectivity.
Fig.\ref{Fig5bis} presents different pairs of order parameters. It is seen that the pair of global parameters (top panel) are more informative and confirm the conclusions drawn by the analysis
of Fig.\ref{Fig5}, i.e. the  structures of the polymorphs approach the one of the Bcc structure at the same temperature ($T=0.7$).

Further support to the conclusion that fully-flexible chains crystallize in a Bcc-like structure is provided by the angular distribution of the monomers belonging to specific shells of neighbors
surrounding a central monomer. To this aim, we modify the Steinhardt procedure and consider the angle $\alpha_{jk}$ between ${\bf r}_{ij}$ and ${\bf r}_{ik}$ where the vector ${\bf r}_{ij}$ joins 
the $i$-th central monomer with the $j$-th one which is $r_{ij}$ apart. Our quantity of interest is the angular distribution function $ADF( \cos \alpha_{jk})$ of the monomers with
$r_{min} \le r_{ij}, r_{ik} \le r_{max}$.
Fig.\ref{Fig6} shows the ADF distribution for the polymorphs and the liquids restricted to the first and the second shells, respectively. As a reference, the ADF distributions of Bcc, Fcc and Hcp crystals
at $T=0.7$ are also plotted. On the basis of the radial distribution function in  Fig.\ref{Fig2} (top), the boundaries of the shells are taken as: $r_{min}=0.8$, $r_{max}=1.35$ (first shell) and
$r_{min}=1.35$, $r_{max}=2.2$ (second shell). It is seen that the ADF distributions of the different polymorphs are distinct. They differ from both the liquid and the reference atomic
crystals in the first shell. Differently, the ADF distributions of the different polymorphs are in excellent agreement with the Bcc ADF in the {\it second} shell. It is worth noting that the ADF of the liquid is nearly flat in the second shell
for $\cos\alpha<0$, signaling the large loss of anisotropy beyond the first shell.

\section{Conclusions}
\label{conclusions}
The isothermal, isobaric spontaneous crystallization of a supercooled polymer melt is investigated by MD simulation of an ensemble of fully-flexible linear chains. Frustration is introduced  
via two incommensurate  length scales, i.e. the bond length and the minimum of the non-bonding potential,
resulting in marked  polymorphism with considerable bond disorder, distortions of both the local packing and the global monomer arrangements.
The crystallization process involves monomer displacements as large as a few diameters. Jump processes are detected in the polymorphs.
The analyses in terms of: i) orientational order parameters characterizing the global and the local order and ii) the angular distribution of the next-nearest neighbors
of a monomer reach the conclusion that the polymorphs are arranged in distorted Bcc-like lattices.

\section*{Acknowledgements}
\label{Ack}
A generous grant of computing time from IT Center, University of Pisa and Dell${}^\circledR$ Italia is gratefully acknowledged.

\end{document}